\documentclass[]{aa}
\usepackage{amsmath,graphicx,amssymb}
\usepackage{txfonts}
\usepackage{natbib}
\usepackage{hyperref}

\newcommand{\kms}{\,km\,s$^{-1}$}
\newcommand{\Teff}{$T_\mathrm{eff}$}
\newcommand{\logg}{$\log g$}

\begin{document}

\title{A Kepler K2 view of subdwarf A-type stars}

\author{G.~M{\"o}senlechner\inst{1}
 \and E.~Paunzen\inst{2}
     \and I.~Pelisoli\inst{3,4}
 \and J.~Seelig\inst{1}
 \and S.~Stidl\inst{1}
     \and H.M.~Maitzen\inst{1}}
\institute{Department of Astrophysics, Vienna University, T{\"u}rkenschanzstraße 17, 1180 Vienna, 
Austria
\and Department of Theoretical Physics and Astrophysics, Masaryk University,
Kotl\'a\v{r}sk\'a 2, 611\,37 Brno, Czech Republic \\
\email{epaunzen@physics.muni.cz}
\and Institut f{\"u}r Physik und Astronomie, Universit{\"a}tsstandort Golm, Karl-Liebknecht-Str. 24/25, 14467 Potsdam, Germany 
\and Department of Physics, University of Warwick, Coventry, CV4 7AL, UK
}

\date{}

\abstract
{The spectroscopic class of subdwarf A-type (sdA) stars has come into focus in recent years
because of their possible link to extremely low-mass white dwarfs, a rare class of objects resulting from binary evolution. Although most sdA stars are consistent with metal-poor halo main-sequence stars, the formation and evolution of a fraction of these stars are still matters of debate.}
{The identification of photometric variability can help to put further constraints on the evolutionary status of sdA stars, in particular through the analysis of pulsations. Moreover, the binary ratio, which can be deduced from eclipsing binaries and ellipsoidal variables, is important as input for stellar models. In order to search for variability due to either binarity or pulsations in objects of the spectroscopic sdA class, we have extracted all available high precision light curves from the Kepler K2 mission.}
{We have performed a thorough time series analysis on all available light curves, employing three different methods. Frequencies with a signal-to-noise ratio higher than four have been used for further analysis.}
{From the 25 targets, 13 turned out to be variables of different kinds (i.e. classical pulsating stars, ellipsoidal and cataclysmic variables, eclipsing binaries, and rotationally induced variables). For the remaining 12 objects, a variability threshold was determined.}
{}

\maketitle

\titlerunning{A Kepler K2 view of sdA stars}
\authorrunning{M{\"o}senlechner et al.}

\begin{table*}[t]
\begin{center}
\tiny
\caption{Target stars for which {\it Kepler K2} data are available.}
\label{stellar_parameters}
\begin{tabular}{clcccccccccc}
\hline
PMF & \multicolumn{1}{c}{EPIC} & $\alpha$ & $\delta$ & $l$ & $b$ & $G$ & \Teff & \logg & $D$ & Var & VSX \\
& & [deg] & [deg] & [deg] & [deg] & [mag] & [K] & & [pc] & & \\
\hline
2917-54556-0089 &     201381939$^{(1)}$     &     181.1339381   &     $-$1.2335514   &     279.3532     &     +59.4739     &     14.800 &     7181(6) &     4.25(3) &     1329(48)     &     N     &          \\
2892-54552-0194 &     201434772     &     180.7233651   &     $-$0.4391811   &     278.0176     &     +60.0633     &     15.618 &     7102(11)     &     6.74(1) &     1917(137)     &     N     &          \\
0515-52051-0079 &     201634373     &     179.3127985   &     +2.5753504    &     272.7818     &     +62.2322     &     15.648 &     7123(10)     &     4.13(6) &     1743(120)     &     N     &          \\
3334-54927-0349 &     202066368     &     102.3520534   &     +17.2566203   &     197.2626     &     +7.3391 &     14.438 &     7583(8) &     4.50(3) &     1799(113)     &     Y     &     *     \\
3334-54927-0089 &     202066487     &     103.3074378   &     +16.3226250   &     198.5148     &     +7.7488 &     14.663 &     7525(8) &     4.50(4) &     2119(126)     &     N     &          \\
2676-54179-0319 &     202066571     &     101.1743567   &     +27.5151115   &     187.4209     &     +10.8037     &     14.677 &     8172(8) &     4.63(4) &     3247(248)     &     Y     &     *     \\
2676-54179-0272 &     202067375     &     101.5634774   &     +27.4023315   &     187.6737f     &     +11.0689     &     14.849 &     8410(8) &     4.38(4) &     4088(489)     &     N     &          \\
2676-54179-0317 &     202067392     &     101.2175806   &     +27.7123640   &     187.2555     &     +10.9220     &     14.878 &     8353(9) &     4.62(4) &     3549(275)     &     Y     &          \\
5293-55953-0322 &     211378898     &     133.8382229   &     +11.3041867   &     216.7222     &     +32.5876     &     18.236 &     --    & -- &   553(50) &     Y     &     *     \\
2428-53801-0439 &     211477347     &     130.7469160   &     +12.8116845   &     213.6377     &     +30.4661     &     15.322 &     8916(12)     &     3.89(1) &     4779(181)     &     Y     &     *     \\
2434-53826-0400 &     211617909     &     135.3063008   &     +14.7846221   &     213.5941     &     +35.3132     &     16.131 &     --    & --    &     1418(90)     &     Y     &     *     \\
2271-53726-0187 &     211823779$^{(2)}$     &     125.0139878   &     +17.6539515   &     206.0801     &     +27.2645     &     15.114 &--   &--   &     2267(139)     &     N     &          \\
1924-53330-0016 &     212003762     &     123.5283466   &     +20.3171373   &     202.7186     &     +26.9310     &     15.765 &--         &--   &     2539(250)     &     Y     &          \\
1927-53321-0533 &     212108396$^{(3)}$ &   126.1970912   &     +22.1869459   &     201.7008     &     +29.9111     &     15.986 &--   &--   &     4268(668)     &     N     &          \\
1929-53349-0356 &     212137838     &     127.2608925   &     +22.7768563   &     201.4298     &     +31.0334     &     15.816 &--         &--         &     2021(235)     &     Y     &          \\
2315-53741-0286 &     212167054     &     126.5574610   &     +23.4182611   &     200.4796     &     +30.6280     &     14.451 &     7745(10)     &     5.06(5) &     2064(105)     &     Y     &     *     \\
2315-53741-0014 &     212168575     &     128.2866706   &     +23.4526826   &     201.0318     &     +32.1433     &     13.935 &     7732(6) &     4.46(4) &     3372(161)     &     Y     &     *     \\
2716-54628-0638 &     212682624     &     211.1947108   &     $-$0.6814277   &     331.8799     &     +50.0228     &     15.729 &     8699(7) &     4.65(3) &     2798(276)     &     Y     &          \\
0696-52209-0410 &     220227479     &     20.7425378    &     +0.9711746    &     139.3066     &     $-$60.8941    &     15.403 &     10279(50)     &     3.79(8) &     4837(755)     &     N     &          \\
4550-55894-0172 &     220489294     &     16.5351818    &     +6.5231553    &     129.4998     &     $-$56.1610    &     19.102 &     10255(60)     &     8.57(7) &     324(11) &     N     &          \\
3779-55222-0306 &     228960916     &     192.7447201   &     $-$2.2001834   &     302.6978     &     +60.6714     &     17.944 &     9725   &     4.17   &     126(2) &     N     &          \\
0381-51811-0200 &     246429212     &     347.2976307   &     $-$0.0761300   &     76.4812 &     $-$53.4638    &     16.864 &     9224(16)     &     4.52(3) &     2098(280)     &     Y     &          \\
3834-56014-0026 &     248407521     &     162.3027266   &     $-$0.1487126   &     250.7756     &     +49.9253     &     18.211 &     6948(35)     &     7.57(9) &     83(1)  &     N     &          \\
5357-55956-0482 &     248833732     &     163.6465341   &     +11.4972242   &     236.8992     &     +58.4400     &     18.590 &     9251(45)     &     8.43(7) &     138(3) &     N     &          \\
2304-53762-0140 &     251353301     &     139.6565309   &     +21.3558369   &     207.5907     &     +41.4934     &     14.436 &     7423(8) &     4.76(4) &     1784(92)     &     Y     &     *     \\
\hline
\end{tabular}
\tablefoot{
The coordinates and astrophysical
parameters were taken from \citet{2019MNRAS.482.3831P} and updated for known WDs. The uncertainties quoted by \citet{2019MNRAS.482.3831P} are also quoted here, though they are only formal fitting errors. The systematic uncertainties are much larger and amount to 5\% in $T_\textrm{eff}$ and up to 1.0~dex in $\log~g$ \citep{2018MNRAS.478..867P}. For EPIC 211617909 and EPIC 211378898, the spectra are dominated by disk emission; therefore, the parameters derived by \citet{2019MNRAS.482.3831P} are unreliable and have been omitted. The photo-geometric distances are taken from \citet{2021AJ....161..147B} on the basis of the {\it Gaia} Early Data Release 3. `PMF' corresponds to the plate-MJD fibre of their SDSS DR12 spectra. Values of \Teff and \logg\ assume solar metallicity and hence should be interpreted as only indicative of the evolutionary class. The last two columns denote
if a star was found variable (Y) or non-variable (N) and if it is known as a variable in the VSX (*). \\
(1) Classified as an extremely metal-poor star by \citet{2017ApJS..228...19C}. 
(2) Previously found to be variable by \citet{2018MNRAS.474.5186R}; we find no variability in K2 data due to the longer cadence.
(3) Classified as `sdB+MS' by \citet{2017A&A...600A..50G}; marginal peaks suggesting possible 
short-term variability found by \citet{2017ApJ...845..171B}.}
\end{center}
\end{table*}

\section{Introduction} \label{introduction}

White dwarfs (WDs) are the final evolutionary state of stars with initial masses of less than about 9\,M$_{\odot}$ \citep{2018MNRAS.480.1547L}. For the evolution 
of single stars, the minimum mass of a WD is around 0.40\,M$_{\odot}$ \citep{2007ApJ...671..761K}. Stars on the main sequence with lower masses have an evolution
time greater than the age of the Universe. Considering the mass-radius relation of WDs, such masses correspond to a minimal \logg\ of
about 6.5. 
On the other hand, the maximum \logg\ of main-sequence A-type stars is about 4.75 in the case of low metallicities. This limit is even lower for higher metallicity because of its radius dependence \citep[e.g.][]{2015MNRAS.450.3708R}. 

Binary evolution can lead to objects having \logg\ values between 6.5 and 4.75 \citep{2016A&A...595A..35I,2018MNRAS.473..693V}. One scenario is
that binary interaction strips away the star’s outer layers during core He burning, resulting in a hot, lower-mass object 
known as a hot subdwarf \citep[see][for a thorough review on these objects]{2016PASP..128h2001H}. 
This mechanism works only for a
certain mass regime because, for stars with masses lower than 2\,M$_{\odot}$, the temperature for burning He is only reached after they become 
degenerate \citep{2004cmpe.conf...31L}. 
As a consequence, if the outer layers of a low-mass progenitor are stripped away before the He burning starts, a degenerate He core with a hydrogen atmosphere 
will be left: a WD. Because the mass of the WDs that result from this mechanism can be much lower than the single star evolution limit, they are known as 
extremely low-mass WDs \citep[ELMs;][]{2020ApJ...889...49B}, which have $4.5 < \log~g < 7.0$. Their precursors, pre-ELMs, can have even lower values of \logg\ before they reach the cooling track. Due to rotational mixing, they can also show atmospheric metals \citep{2016A&A...595A..35I}. Hence, the spectra of (pre-)ELMs can resemble the spectra
of low-metallicity main-sequence A-type stars very closely.

The result is that the evolutionary status of objects that show metal-poor hydrogen-dominated spectra with \logg\ in the range 4--6 \citep[taking the large uncertainties due to line blanketing into account;][]{2017ApJ...839...23B,2018MNRAS.475.2480P} cannot be determined from spectroscopy alone. These objects are grouped into the spectroscopy class of subdwarf A-type (sdA) stars, given that they are located below the canonical (solar metallicity) main sequence \citep{2016MNRAS.455.3413K}. In a series of papers \citep{2019MNRAS.482.3831P}, 
we have attempted to shed light on the evolutionary nature of stars within the sdA spectral class. Recently, \citet{2019ApJ...885...20Y} have shown, using population synthesis models, that most sdA stars are consistent with metal-poor main-sequence stars, in agreement with previous works. The fraction of (pre-)ELMs can vary from 0.1\% to 20\%, depending on the age of the population. 

In summary, the physical properties of sdA stars are basically consistent with four different scenarios: (a) pre-ELMs or ELMs;
(b) blue-stragglers; (c) metal-poor late-type main-sequence stars; and
(d) hot subdwarfs plus main-sequence F, G, K binaries \citep{2018MNRAS.475.2480P}.

Investigating the properties of the light curves of sdA stars offers an independent probe into their evolutionary 
origin \citep[e.g.][]{2018A&A...617A...6B}. Depending on the evolutionary status of the sdA stars, they can be located, for example, within the classical 
pulsational instability strip. Our main aim is to study the (non-)variability of this 
star group in order to shed more light on the evolutionary status of the individual objects.

The long, ultra-precise, and well-sampled time series data provided by space missions, such as the the {\it CoRoT} \citep{Corot}, 
{\it Kepler} \citep{Kepler}, or {\it TESS} \citep{Tess} satellites, have revolutionized the field of variable star research and, in particular, 
asteroseismology. The detection limit for variability signals has decreased dramatically, which has enabled the first 
detailed investigations and classifications of variable stars with very low amplitudes and has revealed unprecedented 
detail and complexity in their light curves. 

In this paper we present the time series analysis of 25 stars classified as sdA that were observed by the {\it Kepler K2} mission. We also derive upper limits of non-variability for almost half the objects.

\section{Target selection, used data, and data reduction} \label{target_selection}

We used the list of 3891 sdA stars previously identified in the Sloan Digital Sky Survey (SDSS) with available and reliable {\it Gaia} Data Release 2 (DR2) astrometric data 
taken from \citet{2019MNRAS.482.3831P}. This is the best currently available dataset for this spectroscopic class. As the next step, we searched
for matches in the complete dataset of the {\it Kepler K2} mission. 
Kepler was launched in March of 2009; it consists of a 0.95\,m Schmidt telescope feeding a 94.6 million pixel 42 charge-coupled device (CCD) detector array 
and covers a wavelength of 4300\AA\ to 8400\AA. 
The {\it Kepler} K2 mission began in November of 2013 and continued until 2019; during this period, {\it Kepler} 
was repurposed for a pointed survey of predetermined targets along the ecliptic plane, in contrast to the original fixed Kepler Input Catalog (KIC) stars. 
K2 typically saved target pixel files (TPFs) of chosen targets and would only download full-frame images (FFIs) twice per campaign. K2 data 
products were mostly long-cadence TPFs (29.4\,min), short-cadence TPFs (1\,min), and long-cadence light curves. Each 
field (115.6 deg$^2$) was observed for approximately 80\,d at a time. K2 also boasted a photometric precision of $\approx$ 300 parts per 
million and a pointing stability of approximately 0.66 pixels. 
K2 observed targets of brightness up to $V$\,=\,4\,mag and down to as low as $V$\,=\,20\,mag with a photometric precision of 10\% (at 30 minutes). 
The resulting K2 data were saved in the K2 Ecliptic Plane Input Catalog (EPIC), and the light curves used in this paper were all retrieved from this catalogue.
In total, there are about 490\,000 light curves available. We found 32 light curves for 25 objects, which are listed in Table \ref{stellar_parameters}. 

For the time series analysis of the light curves, the original and pre-reduced \citep{2014PASP..126..948V} data were used.
The latter take into account that the reduced telescope’s ability to point precisely for extended periods of time has influenced
the photometric precision. The reduction technique accounts for the non-uniform pixel response function of the Kepler detectors by 
correlating flux measurements with the spacecraft’s pointing and removing the dependence. Although there are still artefacts visible,
the quality of the light curve and thus the noise level of the data are significantly improved.

All light curves were examined in more detail using the program package {\sc Period04} \citep{2005CoAst.146...53L}, which performs a discrete Fourier 
transformation. The results from {\sc Period04} were checked with {\sc cleanest} and phase dispersion minimization (PDM) algorithms as 
implemented in the program package {\sc Peranso} \citep{2016AN....337..239P}. The same results were obtained within the derived errors, which depend on 
the time series characteristics (i.e. the distribution of the measurements over time and photon noise). If more than one light curve
was available, we checked both of them for the significant frequencies. For all cases, all the significant frequencies were detected in both
datasets. This provides confidence that no spurious frequencies or artefacts are present in the analysed light curves.

We also merged the light curves of one object for two or more campaigns and searched for long-term trends. However, this task
is extremely restrained by instrumental effects. \citet{2017ApJ...851..116M} described their efforts, starting with the FFIs
of the original {\it Kepler} mission. We were not able to gain any statistically significant improvements or find convincing long-term 
trends. Because the time basis of most light curves is about 30 days, we therefore confined our time series analysis to periods
of 30\,d.

\begin{table*}
\begin{center}
\caption{The results of the time series analysis for all target stars.
The errors in the final digits of the corresponding quantity are given in parentheses.}
\label{tsa_targets}
\footnotesize
\begin{tabular}{cccccccccc}
\hline
PMF & EPIC & Campaign & Frequency & Amplitude & SNR & Type & \multicolumn{3}{c}{Upper Limit}\\
& & & & & & & 0.03 -- 0.5 & 0.5 -- 2.0 & 2.0 -- 25 \\
& & & [c/d] & [Flux] & & & [c/d] & [c/d] & [c/d] \\         
\hline
2917-54556-0089 &     201381939     & C102 & & & & & 0.30 & 0.15 & 0.09 \\
2892-54552-0194 &     201434772     & C102 & & & & & 0.50 & 0.25 & 0.18 \\
0515-52051-0079 &     201634373     & C01 & & & & & 0.18 & 0.12 & 0.12 \\
3334-54927-0349 &     202066368     & C00 & 0.61558(9)& 0.0105(3) & 12.3 & GDOR & & & \\
& & & 0.6594(5) & 0.0094(3) & 11.1 & & & \\
& & & 0.5808(1) & 0.0080(3) & 9.4 & & & \\
& & & 0.5475(7) & 0.0048(3) & 5.7 & & & \\
& & & 1.272(1) & 0.0038(3) & 4.9 & & & \\
3334-54927-0089 &     202066487     & C00 & & & & & 0.18 & 0.18 & 0.15 \\    
2676-54179-0319 &     202066571     & C00 & 0.5594(2) & 0.00525(6) & 38.0 & ACV & & & \\
& & & 0.2826(5) & 0.00202(6) & 14.4 & & & \\
& & & 0.3216(6) & 0.00160(6) & 11.4 & & & \\
& & & 0.842(1) & 0.00092(6) & 6.7 & & & \\
2676-54179-0272 &     202067375     & C00 & & & & & 0.14 & 0.14 & 0.14 \\
2676-54179-0317 &     202067392     & C00 & 15.622(2) & 0.00036(4) & 7.7 & DSCT & & & \\
& & & 16.007(2) & 0.00027(4) & 5.6 & & & \\
& & & 14.317(3) & 0.00026(4) & 5.4 & & & \\
& & & 21.404(3) & 0.00025(4) & 4.9 & & & \\
& & & 15.150(3) & 0.00021(4) & 4.8 & & & \\
5293-55953-0322 &     211378898     & C05, C18 & 15.3052(6) & 0.023(2) & 14.7 & UGSU & & & \\
2428-53801-0439 &     211477347     & C16, C18 & 0.6093(1) & 0.28(1) & & EA & & & \\ 
2434-53826-0400 &     211617909     & C05, C18 & 6.8339(2) & 0.0125(4) & 32.8 & CV & & \\
& & & 13.6685(6) & 0.0052(4) & 16.8 & & & \\
& & & 20.5027(8) & 0.0037(4) & 14.4 & & & \\
2271-53726-0187 &     211823779     & C05, C18 & & & & & 0.23 & 0.19 & 0.09 \\     
1924-53330-0016 &     212003762     & C18 & 0.6476(4) & 0.00194(6) & 13.9 & ROT & & & \\
& & & 0.6227(5) & 0.00143(6) & 10.2 & & & \\ 
& & & 1.312(1) & 0.00058(6) & 4.7 & & & \\    
1927-53321-0533 &     212108396     & C05, C18 & & & & & 0.71 & 0.33 & 0.33 \\
1929-53349-0356 &     212137838     & C05 & 0.22416(4) & 0.00554(3) & 56.3 & ROT & & & \\
& & & 0.20862(9) & 0.00279(3) & 28.3 & & & \\ 
& & & 0.2414(3) & 0.00091(3) & 9.3 & & & \\
& & & 0.4569(3) & 0.00073(3) & 7.7 & & & \\
& & & 0.4373(4) & 0.00069(3) & 7.2 & & & \\   
2315-53741-0286 &     212167054     & C18 & 1.4501(4) & 0.27(1) & & EA & & & \\     
2315-53741-0014 &     212168575     & C05, C18 & 3.0421(2) & 0.126(2) & 26.4 & RRc & & & \\
& & & 6.085(1) & 0.017(3) & 7.4 & & & \\
& & & 9.127(2) & 0.08(3) & 5.8 & & & \\
2716-54628-0638 &     212682624     & C06 & 1.4500(2) & 0.00122(5) & 21.0 & ELL & & & \\
& & & 3.309(1) & 0.00025(5) & 4.9 & & & \\
0696-52209-0410 &     220227479     & C08 & & & & & 0.33 & 0.18 & 0.15 \\
4550-55894-0172 &     220489294     & C08 & & & & & 2.80 & 1.35 & 1.70 \\    
3779-55222-0306 &     228960916     & C102 & & & & & 1.80 & 2.10 & 1.90 \\
0381-51811-0200 &     246429212     & C12 & 4.0908(2) & 0.0039(1) & 24.8 & ELL & & & \\
& & & 2.0460(6) & 0.0014(1) & 8.0 & & & \\    
3834-56014-0026 &     248407521     & C14 & & & & & 1.20 & 0.80 & 0.90 \\    
5357-55956-0482 &     248833732     & C14 & & & & & 2.00 & 1.80 & 2.00 \\
2304-53762-0140 &     251353301     & C16 & 5.94660(5) & 0.0881(8) & 27.5 & EW & & & \\
& & & 2.9724(7) & 0.0076(8) & 4.7 & & & \\
& & & 8.9210(8) & 0.0071(8) & 5.5 & & & \\
& & & 17.840(1) & 0.0052(8) & 7.1 & & & \\
& & & 11.894(1) & 0.0052(8) & 6.1 & & & \\
& & & 23.787(1) & 0.0046(8) & 8.4 & & & \\
\hline
\end{tabular}
\tablefoot{
The determined frequencies, amplitudes, and signal-to-noise ratios for the variable ones are listed in
Cols. 4 to 6. 
For the eclipsing binaries EPIC 211477347 and 212167054, we list the mean depths of the primary minimum.
The next column denotes the type of variability according to the abbreviations of the VSX (see notes below).
The last three columns list the upper limits in different
frequency domains. The absolute numbers depend on the time basis, apparent magnitude,
and the observed sector. It should be noted that harmonics of frequencies were also detected. 
\\
GDOR stands for $\gamma$ Doradus star. ACV stands for $\alpha^2$ Canum Venaticorum variable. DSCT stands for the variable of the $\delta$ Scuti type. UGSU stands for U Geminorum-type variable, quite often called dwarf novae. EA stands for detached eclipsing binary. CV stands for cataclysmic variable. ROT stands for spotted star showing variability due to rotation. RRc stands for RR Lyrae type-c variable. ELL stands for rotating ellipsoidal variable. EW stands for W Ursae Majoris-type eclipsing binary.}
\end{center}
\end{table*}

\begin{figure}
\begin{center}
\includegraphics[width=0.45\textwidth]{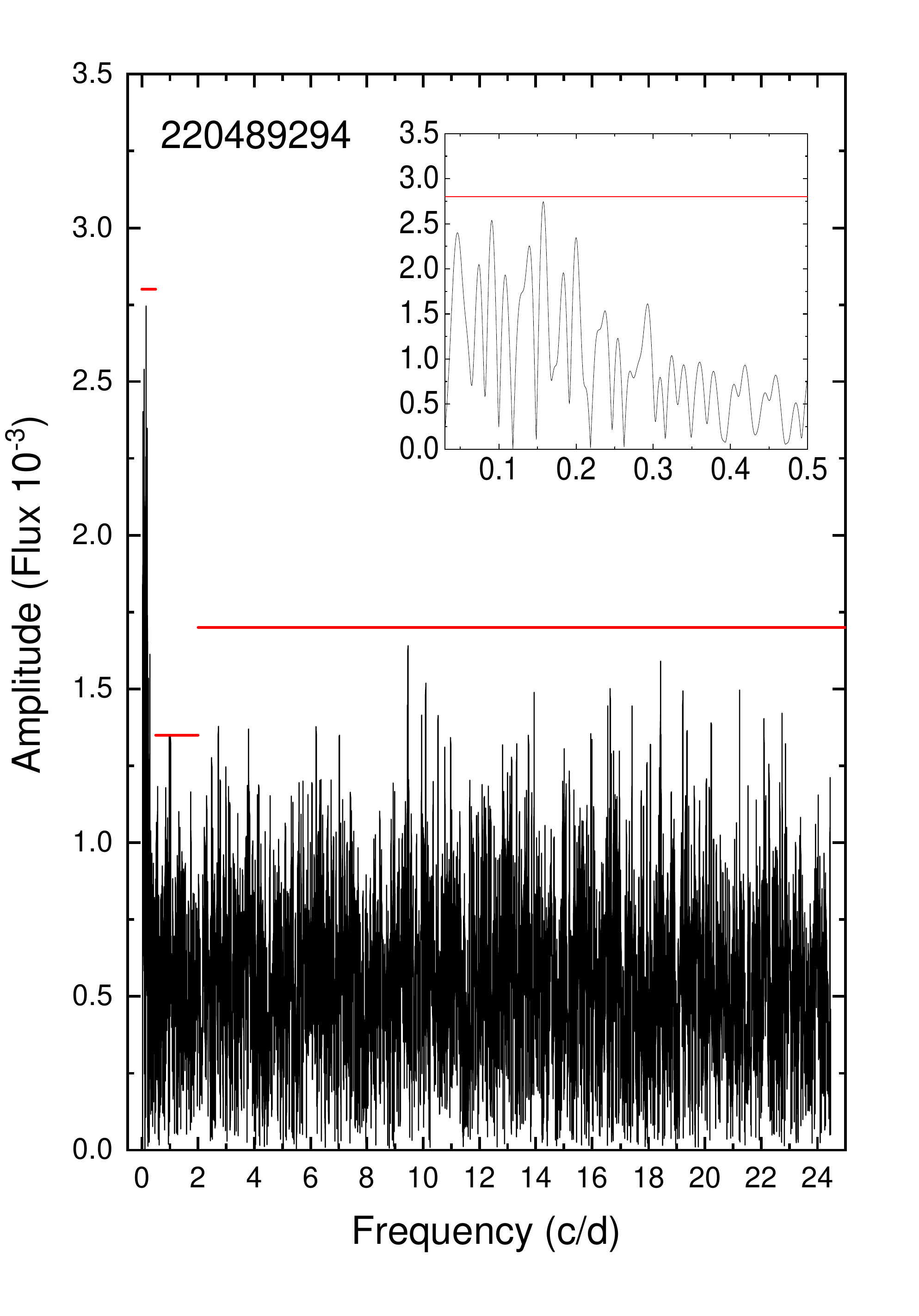}
\caption{Amplitude spectrum of the apparent non-variable star EPIC 220489294, showing how the variability
threshold is defined.}
\label{non_variable}
\end{center}
\end{figure}

Defining the variability threshold is not straightforward. In general, 
the statistical significance of the noise in the Fourier spectrum is underestimated. We employed a conservative 
approach and defined the 
upper limit of variability as the upper envelope of the peaks in an amplitude spectrum. Because the noise levels
change over the frequency range, we derived the values for three regions: 0.03 -- 0.5 c/d, 0.5 -- 2.0 c/d, and 2.0 -- 25 c/d.
One example is shown in Fig. \ref{non_variable}. The red lines show the deduced limits of (non-)variability for the light curve
of EPIC 220489294 (Table \ref{tsa_targets}). 

To check for known variable stars, we applied the version of the International Variable Star Index \citep[VSX;][]{2006SASS...25...47W} available as of 
January 27, 2020 (1\,432\,563 included objects). The search radius for objects was 10''. We detected
eight objects among our sample (see the last column of Table \ref{stellar_parameters}). 

The light curves were analysed together with spectra of the Large Sky Area Multi-Object Fiber Spectroscopic Telescope (LAMOST) DR5 \citep{2019yCat.5164....0L} and the SDSS DR12 \citep{2015ApJS..219...12A} to constrain the evolutionary status of each sdA star. 
The LAMOST
telescope \citep{2012RAA....12.1197C} is a reflecting Schmidt telescope
located at the Xinglong Observatory in Beijing, China. It
boasts an effective aperture of 3.6 to 4.9m and a field of view
of 5$\degr$. Due to its unique design, LAMOST is able to
take 4000 spectra in a single exposure with a spectral resolution
$R \sim 1800$, a limiting magnitude of about 19\,mag in $r$, and a wavelength coverage
from 3700 to 9000\AA. LAMOST is therefore particularly
suited to survey large portions of the sky and is dedicated to a
spectral survey of the entire available northern sky. LAMOST
data products are released to the public in consecutive data releases
and can be accessed via the LAMOST spectral archive\footnote{http://www.lamost.org}. The SDSS uses a dedicated 2.5m wide-field telescope, instrumented with a sequence of sophisticated imagers and spectrographs.
The wavelength coverage of the spectra is from 3800 to 9200\AA\ for
the SDSS spectrograph (up to Plate 3586), and 3650 to
10\,400\AA\ for the BOSS spectrograph, with a resolution of
1500 at 3800\AA\ and 2500 at 9000\AA\ \citep{2011AJ....142...72E}. 

\section{Results} \label{results}

\begin{figure}[t]
\begin{center}
\includegraphics[width=0.45\textwidth]{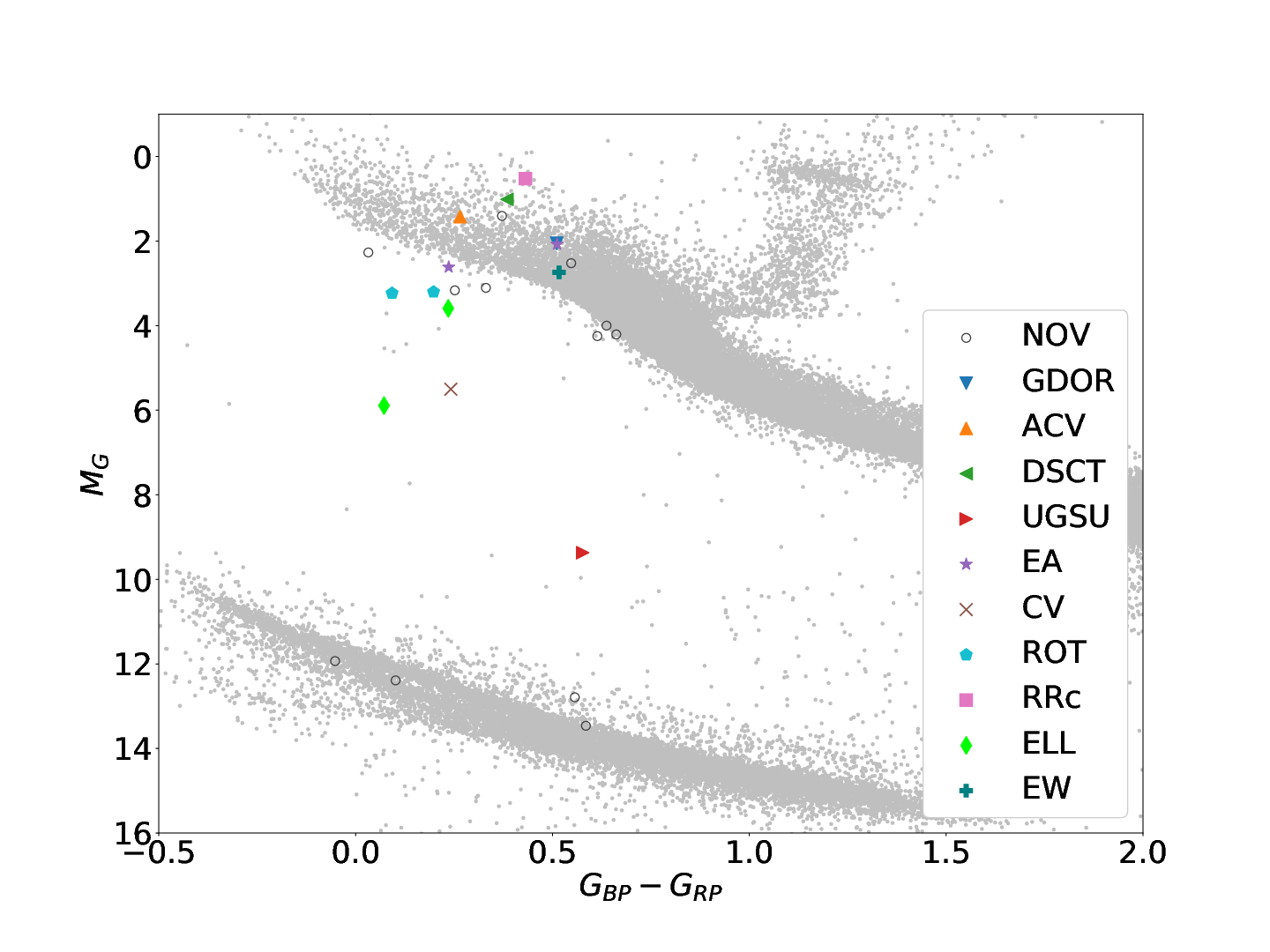}
\caption{HRD, i.e. $M_G$ versus $(BP-RP)$ diagram, of our targets, together with stars from the Sample C presented in
\citet{2018A&A...616A..10G} for guidance. The location of our targets together with their variability type is marked with different
symbols. `NOV' indicates that the target was `not observed to vary'. The remaining acronyms are the same as in Table~\ref{tsa_targets}.}
\label{CMD}
\end{center}
\end{figure}

In Table \ref{tsa_targets} we present all the results of our time series analysis. The detected variable stars among the sdA stars (Table \ref{tsa_targets}) show quite a variety in terms of type (Col. 7 of the table), 
reflecting the different evolutionary status of the spectral sdA class; this is also clear from their different locations in the {\it Gaia} Hertzsprung-Russell diagram (HRD) shown in Fig.~\ref{CMD}. All three stars among the high \logg \ (i.e. the WD domain EPIC 220489294, 248407521, and 248833732) are found to
be constant. The other non-variable objects are distributed along the complete temperature range. The location of variable objects shows good agreement with their observed variability, considering what we know about the pulsation instability strip and the occurrence of $\alpha^{2}$ Canum Venaticorum (ACV) variables, as discussed in more detail below.

\subsection{Astrophysical parameters}

\citet{2019MNRAS.482.3831P}
applied a grid of synthetic spectra to the observed ones from the SDSS in order to derive the
astrophysical parameters. This method was especially optimized for subdwarfs. We have updated this procedure and checked it with the literature for known WDs \citep{2013ApJS..204....5K,2015A&A...583A..86K}.

\begin{figure}[t]
\begin{center}
\includegraphics[width=0.45\textwidth]{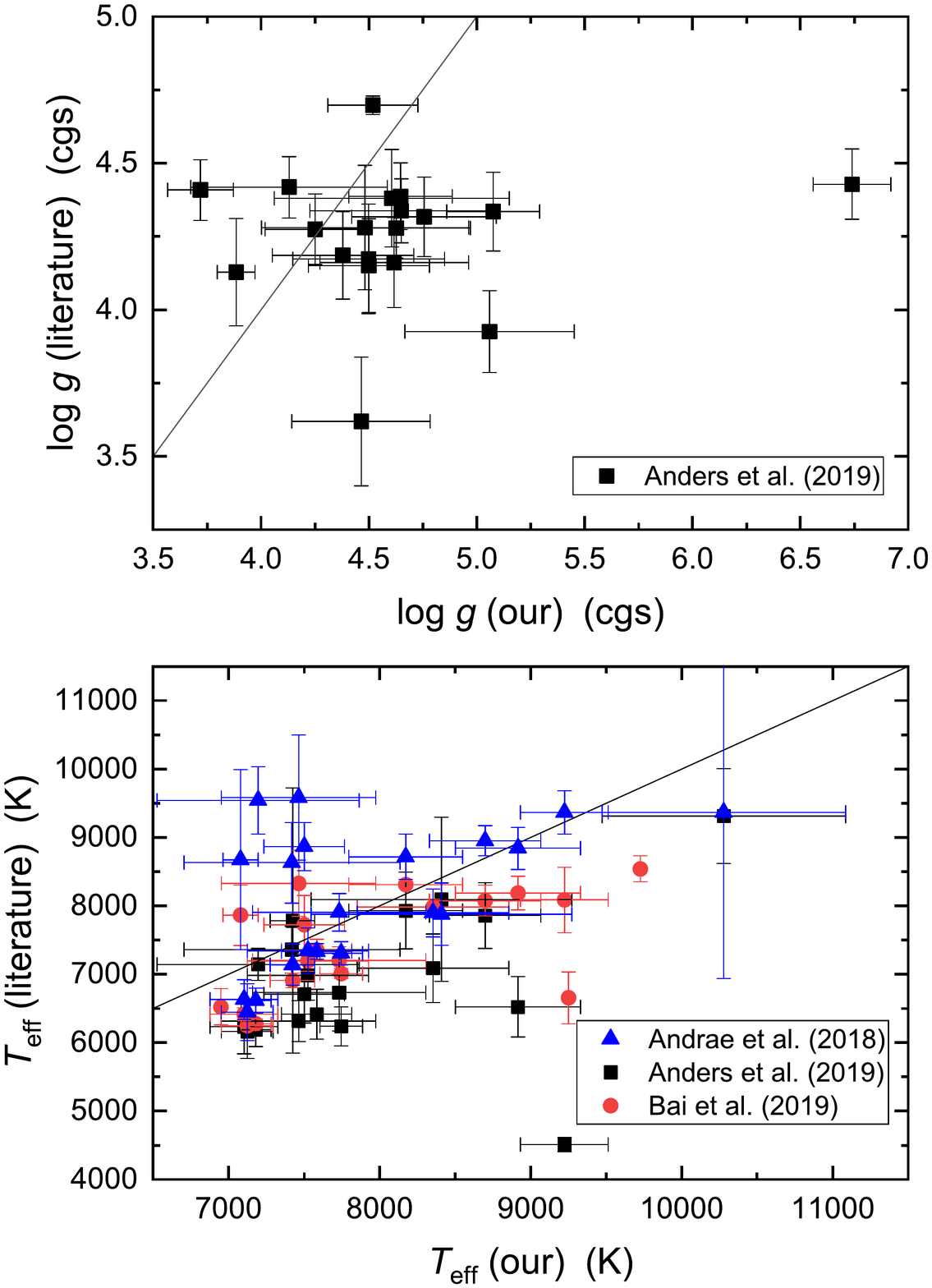}
\caption{Comparison of the \Teff\ (lower panel) and \logg\ (upper panel) values with those from the
literature. The \logg\ values were derived by automatic pipelines
\citep{2018A&A...616A...8A,2019A&A...628A..94A,2019AJ....158...93B}.}
\label{comparison_literature}
\end{center}
\end{figure}

We compared the astrophysical parameters listed in Table \ref{stellar_parameters} with estimates from the literature. For this, we used the values from 
\citet{2018A&A...616A...8A}, \citet{2019A&A...628A..94A}, and \citet{2019AJ....158...93B}, which are all based on an automatic pipeline 
and the {\it Gaia} DR2. In Fig. \ref{comparison_literature} the comparison is presented. 
The \Teff\ values show a 
wide spread around the line of equality. This is also true if we compare the literature values among themselves.
Additionally, due to the rather large uncertainties, most of the data points fall on the identity line in a 2$\sigma$ range.
For the \logg\ values, only \citet{2019A&A...628A..94A} can be used for a comparison. They do not include any WDs
in their study. Again, we see a wide spread, which is to be expected. To get more accurate astrophysical parameters, high resolution 
spectra are needed, which is beyond the scope of this paper.

Furthermore, we checked the astrophysical parameters of the three WDs (i.e. EPIC 
220489294, 248407521, and 248833732). Because they cover a wide variety of spectral types (DB, DA, and DC),
we used different appropriate atmosphere models \citep{2010MmSAI..81..921K} and spectral 
classifications \citep{1982A&A...116..147K}. The estimated \Teff\ and \logg\ values were 
checked with the location in the HRD \citep{2018MNRAS.480.1547L}. No inconsistencies were found.

\subsection{Main-sequence and horizontal branch pulsators}

\subsubsection{EPIC 202066368}

\begin{figure}[t]
\begin{center}
\includegraphics[width=0.45\textwidth]{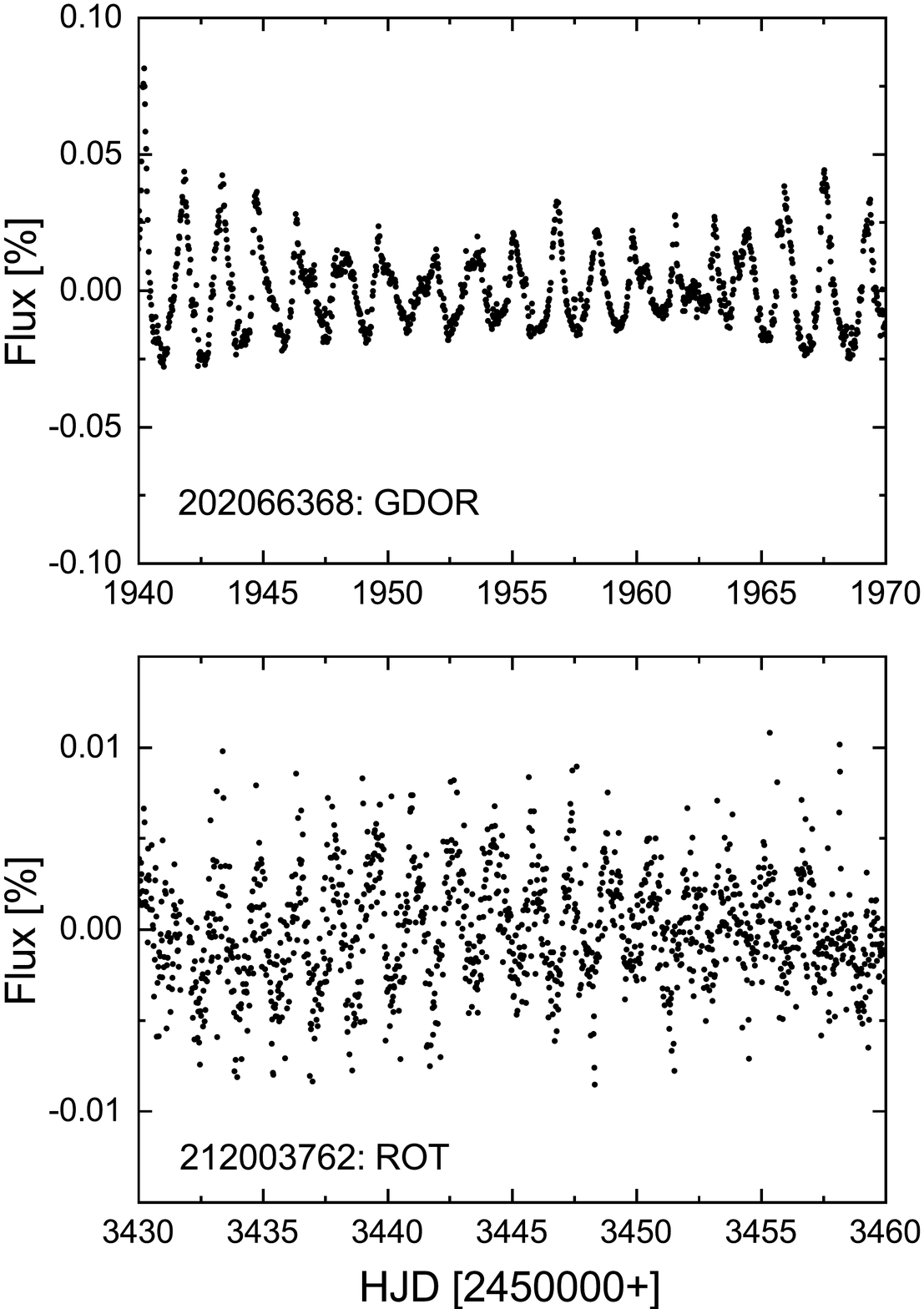}
\caption{Light curves of the pulsating star EPIC 202066368 (upper panel) and the rotational-induced variability of EPIC 212003762 (lower panel).
The multi-periodic variations for EPIC 202066368 are clearly visible.}
\label{GDOR_SPB}
\end{center}
\end{figure}

\citet{2015A&A...579A..19A} classified the light curve of EPIC 202066368 as
quasi-periodic with a period of 1.621151\,d but did not study the source of the variability. We identify the same dominant period as \citet{2015A&A...579A..19A}, as well as four other previously unidentified frequencies. Given the effective temperature, observed eigenfrequencies,
and light curve characteristics, we identify this star as being a $\gamma$ Doradus
pulsator \citep{2011MNRAS.415.3531B}. These variables are characterized by a high-order, 
low-degree, non-radial gravity-mode (g-mode) pulsation \citep{1999PASP..111..840K}, which is thought to 
be driven by the convective flux blocking mechanism \citep{2000ApJ...542L..57G}. They are 
encountered between spectral types A7 and F7, consistent with the physical parameters found for this object (see Table~\ref{stellar_parameters}), although other sources have
shifted the red border of the corresponding instability strip to somewhat hotter temperatures 
(spectral type F5; cf. \citealt{2015pust.book.....C}, and references therein). 
The multi-periodic variations are clearly seen in Fig. \ref{GDOR_SPB}.
We employed the empirical pulsation period versus 
\Teff\ relation derived by \citet{2018NewA...62...70I} for 109 $\gamma$ Doradus stars.
It yields a period of about 0.55\,d for the \Teff\ taken from Table \ref{stellar_parameters}. 
The found periods are between 0.5475 and 0.6594\,d, which perfectly coincidences with the
expected value. EPIC 202066368 is therefore a main-sequence $\gamma$ Doradus pulsator, the first identified 
among sdA stars.

\subsubsection{EPIC 202067392}

\begin{figure}[t]
\begin{center}
\includegraphics[width=0.45\textwidth]{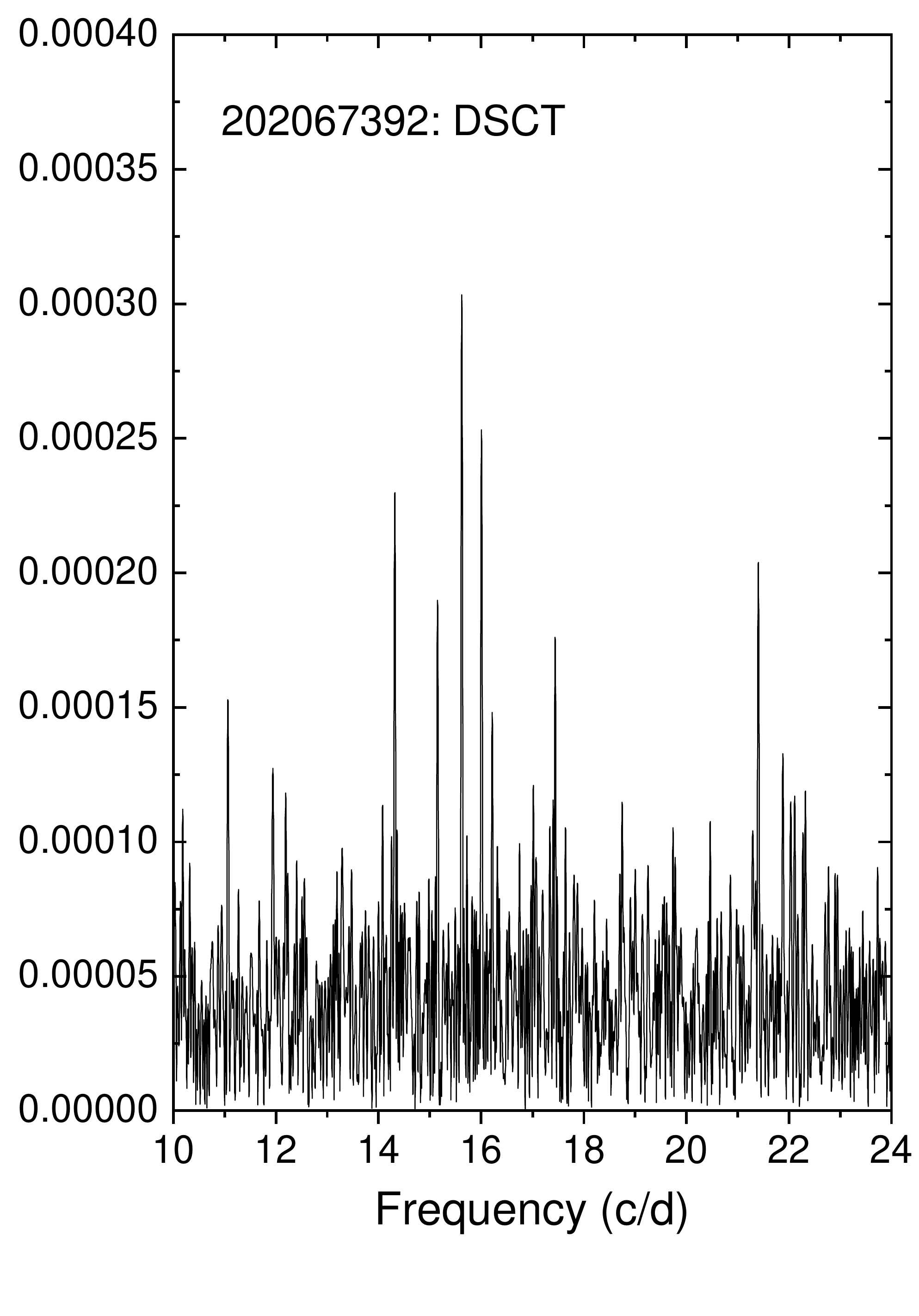}
\caption{Rich amplitude spectrum of the variable of the $\delta$ Scuti-type pulsator EPIC 202067392. The found amplitudes and the noise
level are quite small.}
\label{DSCT}
\end{center}
\end{figure}

This star was included by \citet{2010A&A...522A..88S} in their catalogue of blue horizontal branch stars at a distance of 6.41\,kpc from
the Sun. With the new {\it Gaia} data (Table \ref{stellar_parameters}), we now know that this
object is much closer (about 3.5\,kpc) to the Sun and therefore less luminous than previously thought. Here, we report its variability for the first time.
The light curve and found periods are typical for a $\delta$ Scuti-type pulsation. 
They are multi-periodic pulsators of luminosity classes V to III that boast masses between about 
$1.5 M_{\odot} < M < 4.0 M_{\odot}$ \citep{2010aste.book.....A}. 
The observed light changes are caused by multiple radial and non-radial low-order pressure modes (p modes), 
which are excited through the $\kappa$ mechanism \citep{2009AIPC.1170..403H}. In evolved $\delta$ Scuti stars, 
so-called mixed modes are often observed. These are pulsation modes exhibiting g-mode characteristics 
in the interior and p-mode characteristics near the stellar surface \citep{2016MNRAS.460.1970B}. 
If we take the period luminosity relation from \citet{2019MNRAS.486.4348Z}, the period with the
highest amplitude gives an absolute magnitude of +2.17\,mag. Taking the astrophysical
parameters from Table \ref{stellar_parameters}, we derive a value of about +2\,mag with an error
of about $\pm$0.3\,mag. The main contribution here is the error of the parallax. Because the
detected variability is multi-periodic (Fig. \ref{DSCT}) within the proper frequency range, we conclude that 
EPIC 202067392 is a $\delta$ Scuti-type pulsator. The high gravity found by 
\citet{2019MNRAS.482.3831P} (4.62; see Table~\ref{stellar_parameters}) is at odds with the fact that this 
is likely a young object, hence with high metallicity, in agreement with its position in Fig.~\ref{CMD}. 
This can be explained by the fact that \citet{2019MNRAS.482.3831P} assumed solar metallicity in their 
fits, leading to high systematic uncertainties in \logg.

\subsubsection{EPIC 212168575}

\begin{figure}[t]
\begin{center}
\includegraphics[width=0.45\textwidth]{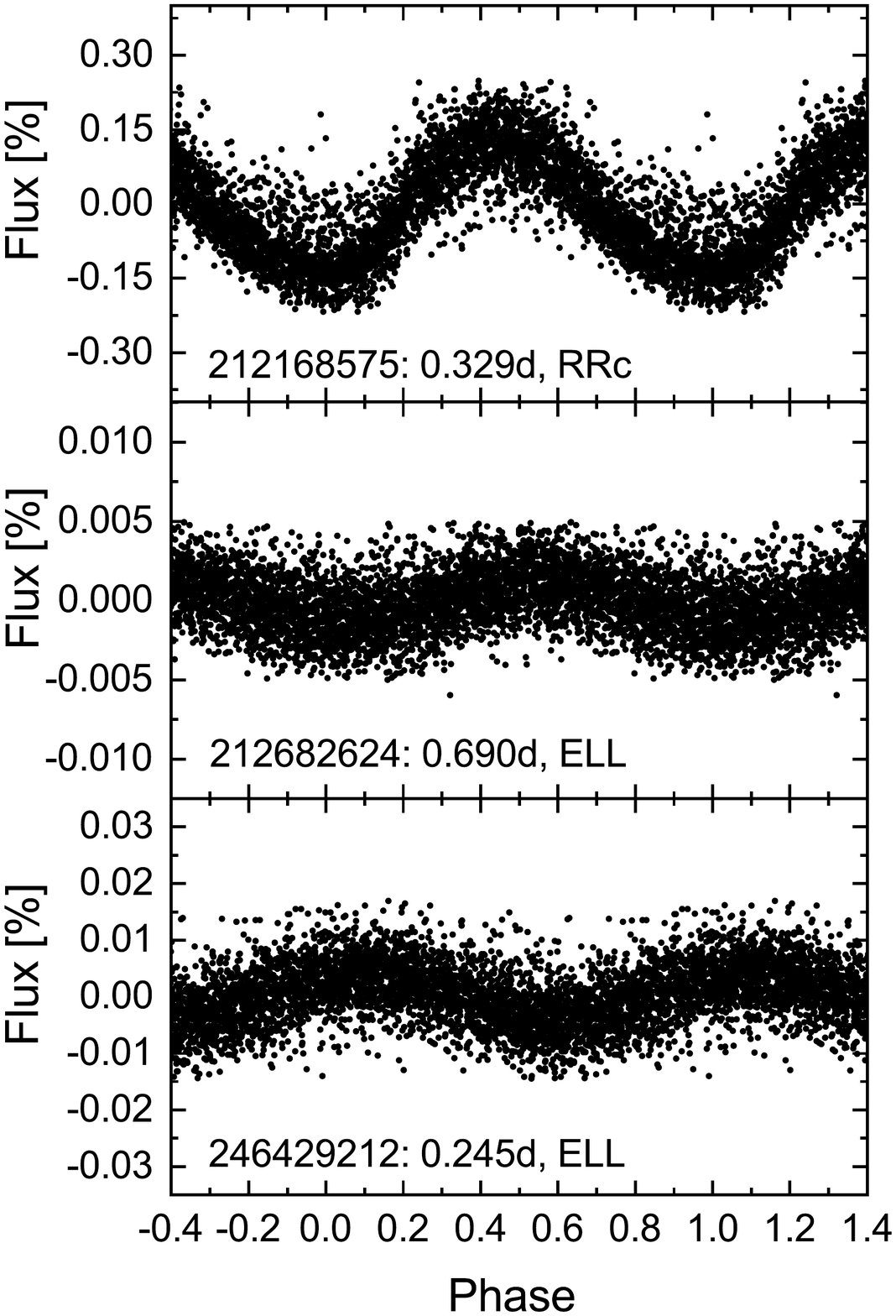}
\caption{Phased light curves of EPIC 212168575 (upper panel), EPIC 212682624 (middle panel), and EPIC 246429212 (lower panel).}
\label{RRc_ELL_var}
\end{center}
\end{figure}

This star has previously been identified as an RR Lyrae star by \citet{2013AJ....146...21S}, using data from the The Lincoln Near-Earth Asteroid Research (LINEAR) survey. We confirm this result with the higher-quality K2 light curve, shown in Fig. \ref{RRc_ELL_var}, which confirms EPIC 212168575 as an RRc variable. Classical RR Lyrae are evolved stars that burn helium in their cores. In the HRD, they are located at the intersection 
of the horizontal branch and the classical instability strip, in which the $\kappa$ mechanism operating in the hydrogen and helium partial 
ionization zones drives the pulsation. RRc stars pulsate in the first overtone radial mode \citep{2019ApJS..244...32P}. It is important to mention that RR Lyrae-like pulsations can also be observed in pre-ELMs: Because of their complicated evolution, which can involve episodes of residual burning, they can have properties within the RR Lyrae instability strip. There is at least one observed system with such properties \citep{2012Natur.484...75P}, with another candidate observed by \citet{2018A&A...617A...6B}. The former could be confirmed as a pre-ELM because it is in an eclipsing system, which allowed its mass to be determined as only 0.26~$M_{\odot}$. We unfortunately detect no eclipses for EPIC 212168575. Considering that \citet{2012Natur.484...75P} estimated that only 0.2\% of RR Lyrae are pre-ELMs, and that only a small fraction of sdA stars are expected to not be main-sequence stars, we conclude that EPIC 212168575 is likely a classical RR Lyrae star. This, combined with the position in Fig.~\ref{stellar_parameters}, suggests that the $\log~g$ estimated by \citet{2019MNRAS.482.3831P} and quoted in Table~\ref{stellar_parameters} is overestimated, illustrating the importance of performing additional analyses other than spectral fitting for sdA stars.

\subsection{A new $\alpha^{2}$ Canum Venaticorum variable}

\begin{figure}[t]
\begin{center}
\includegraphics[width=0.45\textwidth]{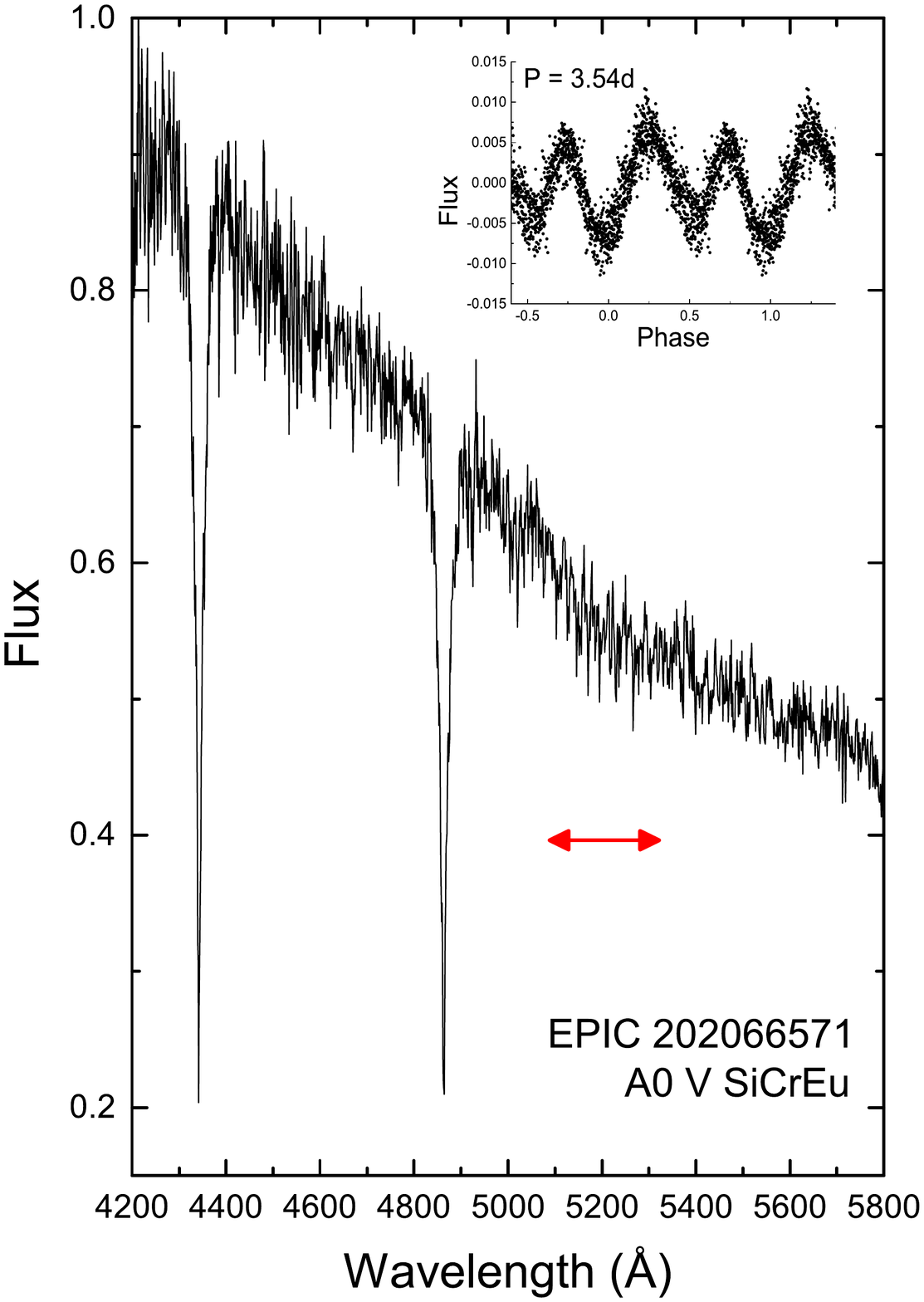}
\caption{Part of the LAMOST spectrum for the bona fide CP star EPIC 202066571.
Marked is the 5200\AA\ flux depression characteristic for members of this star group. The phased light curve is shown in the upper-right corner.}
\label{202066571}
\end{center}
\end{figure}

The light curve of EPIC 202066571, shown in the inset of Fig.~\ref{202066571}, resembles that of a classical ACV variable. These are magnetic, chemically peculiar
(CP) stars of the upper main sequence that show a non-uniform distribution of chemical elements. 
This distribution manifests itself in the formation of spots and patches of enhanced element
abundance \citep{1981A&A...103..244M}, in which flux is redistributed through bound-free and 
bound-bound transitions. Therefore, as the star
rotates (normally below 50\,\kms), strictly periodic changes are observed in the
spectra and brightness of many CP stars, which are
satisfactorily explained by the oblique rotator model \citep{1950MNRAS.110..395S}.
The light curves are stable over several thousand rotational cycles \citep{2015AN....336..981B}.
The found period (about 3.54\,d) is typical for ACV variables. We have to emphasize that this is not
the frequency with the highest amplitude because EPIC 202066571 shows a 
double-wave structure (Fig. \ref{202066571}). Such a behaviour has been observed in several ACV variables \citep{1980A&A....89..230M,2016AJ....152..104H}. 
The classification of the available
LAMOST and SDSS spectra according to appropriate Yerkes (MKK) standard stars \citep{2011AN....332...77P} results
in a spectral type of A0\,V SiCrEu, which is typical for a magnetic CP star. As the parameters from \citet{2019MNRAS.482.3831P} quoted in Table~\ref{stellar_parameters} assumed stellar metallicity, the derived \Teff\ is lower than expected for an A0\,V. Most importantly, 
we detected the typical 5200\,\AA\ flux depression, as shown in Fig. \ref{202066571}.
It was found that this spectral feature occurs solely in these stars and is
correlated with the organized local stellar magnetic field \citep{2007A&A...469.1083K}.
It is well known that the effects of radiative diffusion are
not restricted to the main-sequence domain; they are also predicted in stars with high gravities as well as 
in subdwarfs \citep{2011A&A...529A..60M}. However, a detailed abundance analysis on the basis
of high resolution spectra for EPIC 202066571 is needed to reveal its true nature. This newly identified ACV variable is the first among sdA stars.

\subsection{Two composite hot subdwarfs}

The SDSS and LAMOST spectra for EPIC 212003762 and EPIC 212137838 seem to indicate that they are B-type stars rather than cooler sdA objects. \cite{2016ApJ...818..202L} classified EPIC 212003762 as a composite of a hot subdwarf and a main-sequence star. For EPIC 212137838, \citet{2016MNRAS.457.3396P} concluded from a spectral energy distribution (SED) fitting that it is a 
binary system of a hot component (classification: SdB 0.2 He 9) and a cool component. 
The classification scheme is according to \citet{2013A&A...551A..31D}.

To further investigate the SED of both objects, we employed
the VO Sed Analyzer (VOSA\footnote{http://svo2.cab.inta-csic.es/theory/vosa/} 
v6.0) tool \citep{2008A&A...492..277B} for fitting the available photometry. The contributions of the hot and cool components for each star are visible.
The LAMOST and SDSS spectra confirm this conclusion, clearly showing the \ion{Mg}{I} triplet at 5180\AA\ and the G band. Both objects are included in the hot subdwarf catalogue of \citet{2017A&A...600A..50G} as composite hot subdwarfs. As pointed out in \citet{2018MNRAS.475.2480P}, about 0.5\% of the sdA stars are hot subdwarfs with F,G,K companions. The periods observed here are, however, too long for hot pulsating subdwarfs \citep{2017MNRAS.466.5020H,2018MNRAS.474.4709K}. Assuming the variability comes from the B-type component, these objects could be interpreted as slowly pulsating B-type (SPB) stars. They are main-sequence stars of spectral types B2 to B9, which show non-radial g-mode oscillations driven by the $\kappa$ mechanism acting on the iron bump \citep{1993MNRAS.262..213G}. However, as can be clearly seen in Fig.~\ref{CMD}, where these objects are shown as cyan pentagons, both are under-luminous, suggesting that they are indeed subdwarfs. In this case, the most plausible explanation for the photometric variability, which is reported here for the first time, is that it actually originates in the cooler, unseen companion and is caused by stellar rotation and convection. It results from the change in brightness as spot(s) move in and out of the visible hemisphere of the star. Tens of thousands of rotational variables have been identified by the Kepler \citep[e.g.][]{2013A&A...560A...4R} and K2 missions \citep[e.g.][]{2020A&A...635A..43R}. The light curves often show a second period close to the dominant rotation period, as found here, which is interpreted as surface differential rotation. The relatively short rotation periods detected here are consistent with the formation scenario for hot subdwarfs with cool companions \citep{2002MNRAS.336..449H}: Roche-lobe overflow causes the hot subdwarf progenitor to lose its outer layers, and mass is accreted by the companion, causing it to spin up \citep[see e.g. the $v \sin i$ values in Table 5 of][]{2018MNRAS.473..693V}.

\subsection{Two new ellipsoidal variables}

EPIC 212682624 and EPIC 246429212 show light curves (Fig. \ref{RRc_ELL_var}) consistent with ellipsoidal variables (ELLs), similar to, for example, KIC 4570326 \citep{2014IAUS..301..413G}.
This group of variables consists of close binaries whose components are distorted by their mutual gravitation but whose orbital 
inclinations are too small to show eclipses in the line of sight \citep{1986Ap&SS.125...69B}. The light variations are
a combination of three effects: tidal distortion, reflection, and beaming. Beaming in particular is induced by the stellar radial 
motion, which results in an increase (decrease) in brightness when the star is approaching (receding from) the observer. Due
to these effects and the many free parameters, the light curves of ELLs show a wide variety of shapes and amplitudes 
\citep{2016AcA....66..405S}.

\begin{figure}[t]
\begin{center}
\includegraphics[width=0.45\textwidth]{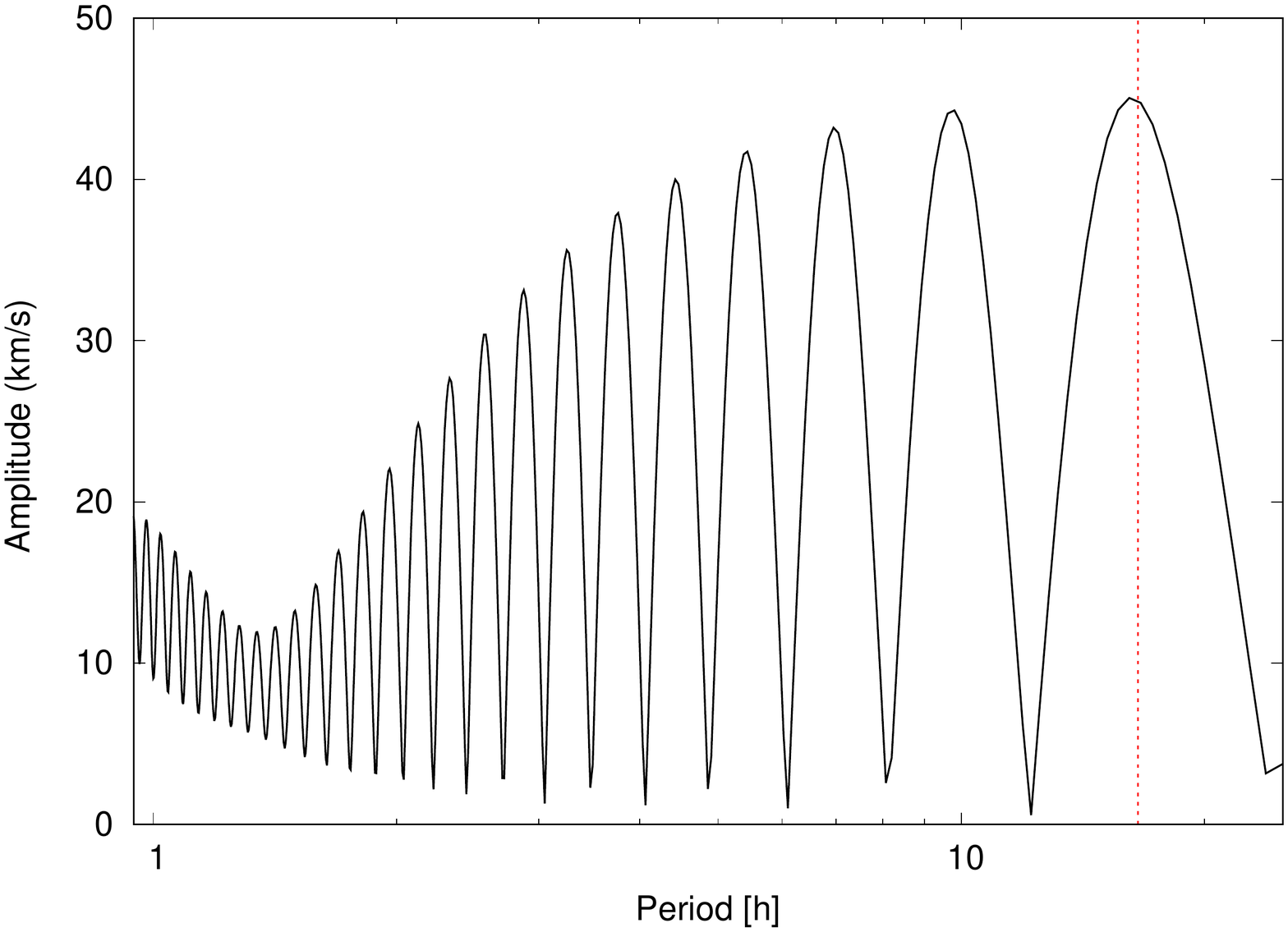}
\caption{Fourier transform for the SDSS radial velocities of EPIC 212682624. The dashed red line marks the photometric period, which is consistent with one of the possible aliases.}
\label{2716_ft}
\end{center}
\end{figure}

\begin{figure}[t]
\begin{center}
\includegraphics[width=0.45\textwidth]{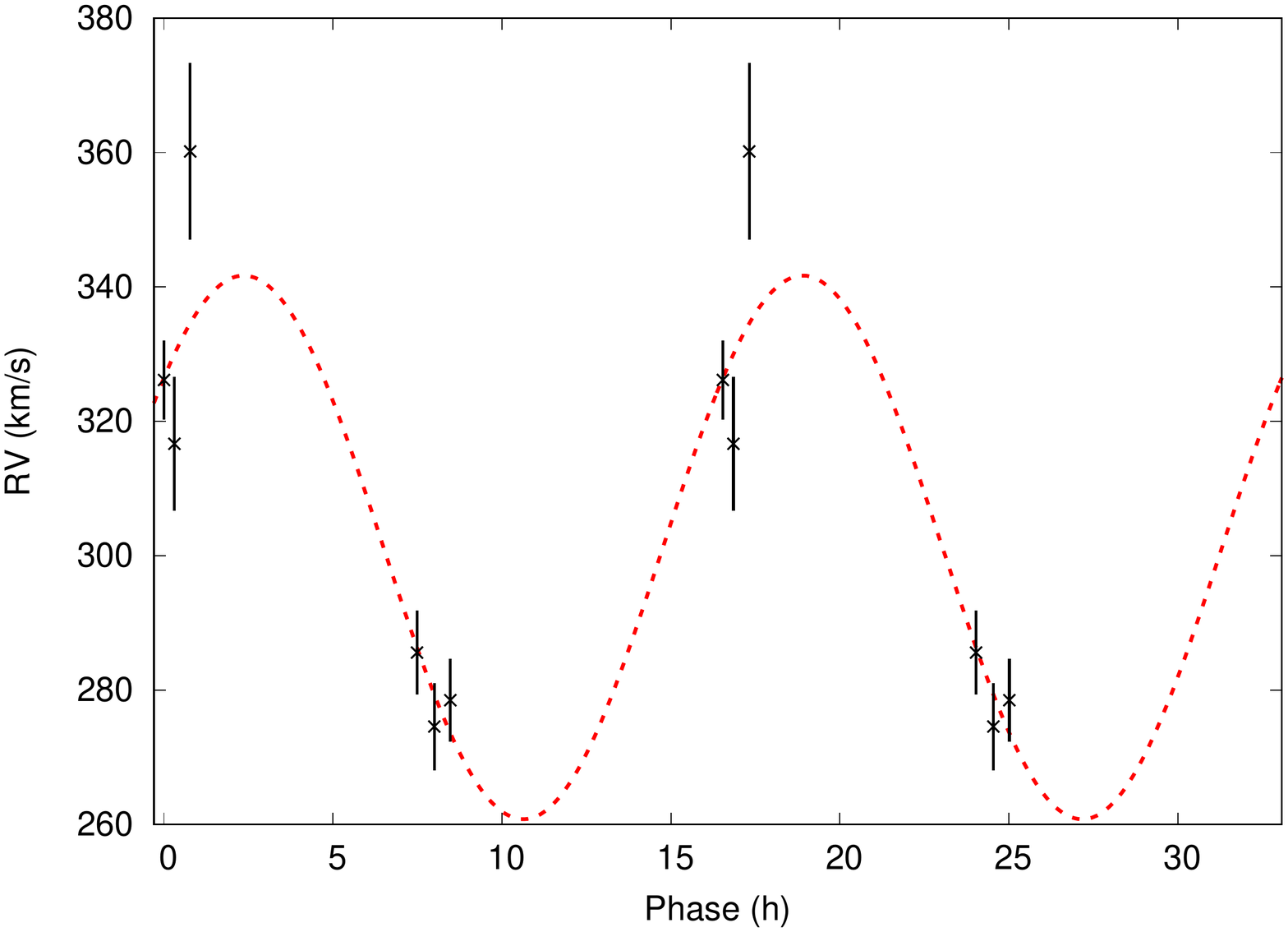}
\caption{SDSS radial velocities of EPIC 212682624, folded to the period in Fig.~\ref{2716_ft} that is consistent with the photometric period. The dashed red line shows a sinusoidal fit, assuming a circular orbit.}
\label{2716_rv}
\end{center}
\end{figure}

A good quality SDSS spectrum is available for EPIC 212682624. It is composed of six exposures taken over 25 hours. We noticed a hint of radial velocity variation (around 50\kms) between these sub-exposures. Using the package {\sc rvsao} \citep{1998PASP..110..934K} and a spectral template with the parameters from Table~\ref{stellar_parameters}, we estimated the radial velocities from each SDSS spectrum. We then calculated the Fourier transform of these velocities, which is shown in Fig.~\ref{2716_ft}. The dashed red line marks the dominant period from the photometry. There are many possible aliases from the radial velocities alone, but one ($1.4509\pm0.2575$~c/d) coincides with the photometric period, providing further evidence for the origin of the photometric variability. Figure~\ref{2716_rv} shows the radial velocities folded to this period, together with a sinusoidal fit, which give an amplitude of $K = 40 \pm 16$~km/s. Given the large uncertainty in the amplitude, the mass function for the unseen companion is not well constrained ($f_2 = 0.0047 \pm 0.0056~M_{\odot}$). Furthermore, this object is located below the main sequence (upper green diamond in Fig.~\ref{CMD}) in a region with relatively low extinction \citep[E(B-V) = 0.0279;][]{2011ApJ...737..103S}. This, combined with the binarity and the ellipsoidal variation, suggests that the object is an ELM with an unseen companion.

\begin{figure}[t]
\begin{center}
\includegraphics[width=0.45\textwidth]{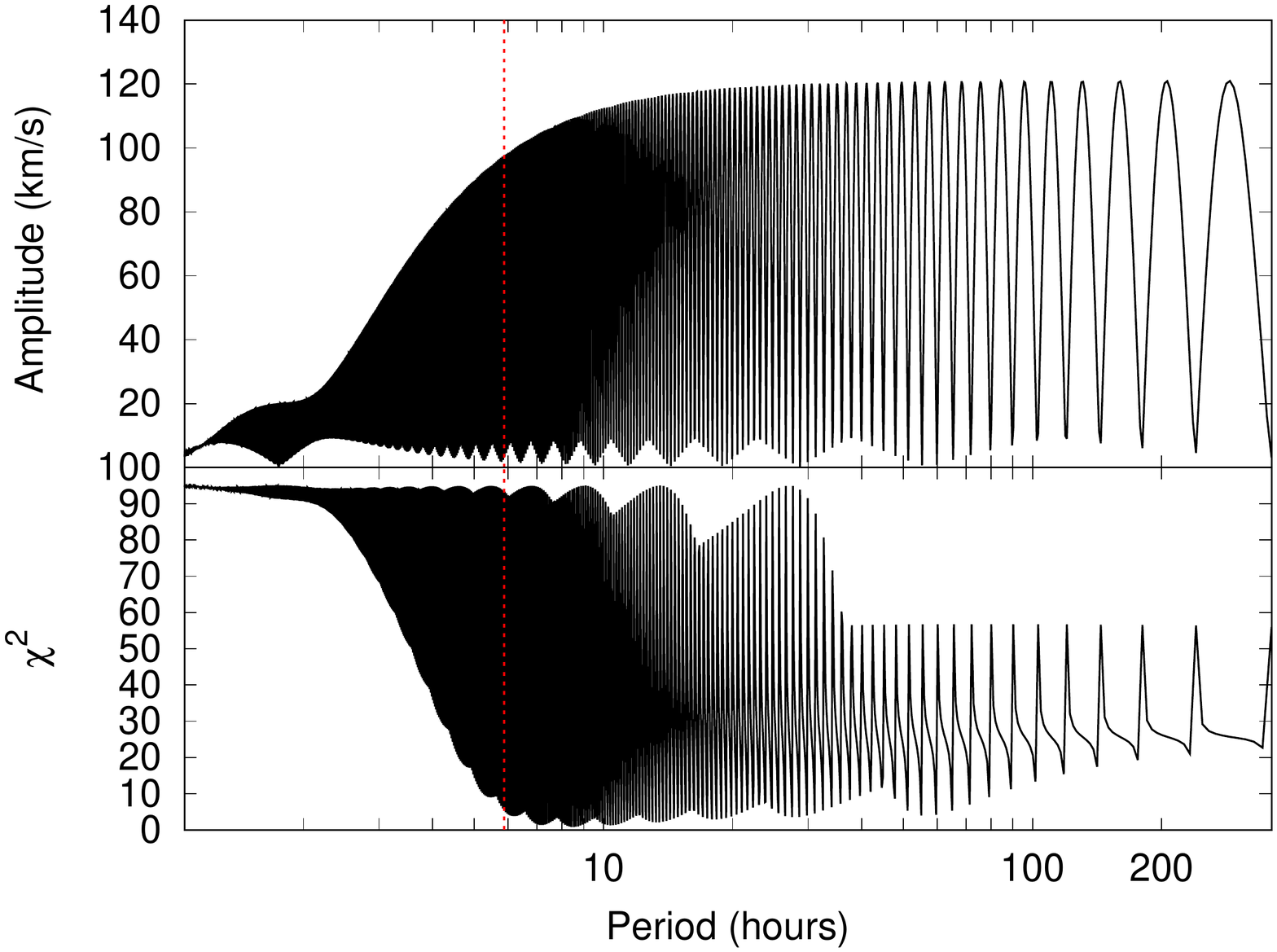}
\caption{Top panel: Fourier transform for the radial velocities of EPIC 246429212, as obtained from SOAR spectra. Bottom panel: Values of $\chi^2$ for orbital solutions calculated at each period, assuming a circular orbit. The dashed red line marks the photometric period. The radial velocities do not constrain the period well, but a solution with low $\chi^2$ is possible near the photometric period.}
\label{0381_ft}
\end{center}
\end{figure}

\begin{figure}[t]
\begin{center}
\includegraphics[width=0.45\textwidth]{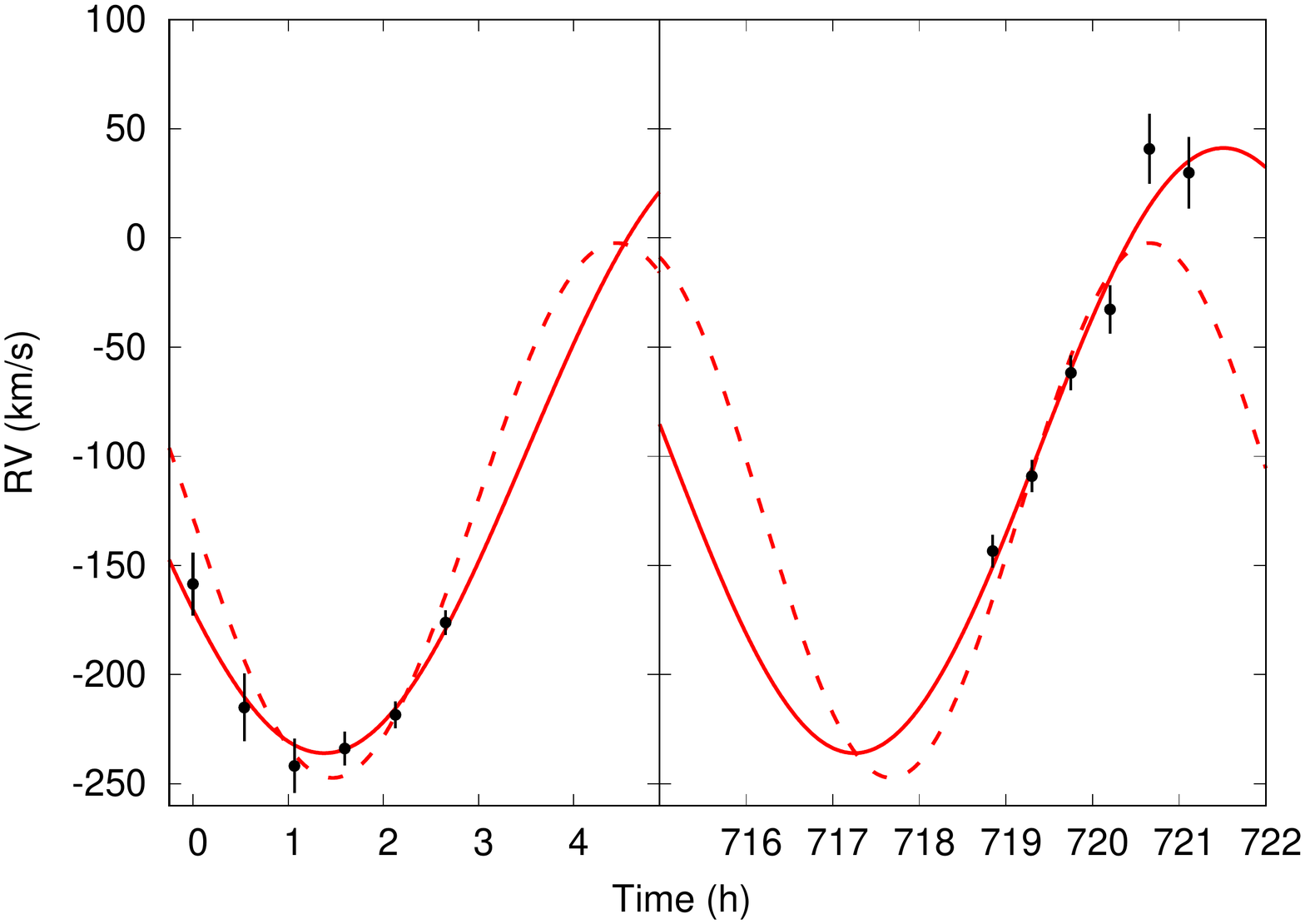}
\caption{Obtained radial velocities for EPIC 246429212 shown together with two possible solutions. The solid line assumes the period with lowest $\chi^2$ in Fig.~\ref{0381_ft}, and for the dashed line we have chosen the alias closest to the photometric period.}
\label{0381_rv}
\end{center}
\end{figure}

EPIC 246429212 was previously selected by \citet{2008ApJ...684.1143X} as a blue horizontal branch star candidate. The phased light curve
is shown in Fig. \ref{RRc_ELL_var}. The short orbital period suggests that the binary cannot harbour a horizontal branch star. The position far below the main sequence in Fig.~\ref{CMD} also supports the idea that the object is instead an ELM. We obtained radial velocities for this star with the Goodman High Throughput Spectrograph at the SOAR 4~m telescope using a 1200~l/mm grating and a 1" slit, obtaining a resolution of $\approx 2$\AA, as part of the programme SO2018B-002 (PI: Pelisoli). Each spectrum was paired with an arc-lamp exposure to guarantee precise radial velocities. The radial velocities were calculated in the same way as for EPIC~212682624. These velocities alone do not constrain the period well -- many aliases are possible, and similar values of $\chi^2$ are obtained assuming a large range of periods (see Fig. \ref{0381_ft}). The best period given by the radial velocities is $2.8\pm0.8$~c/d, but there is also a possible alias at $4.0214\pm0.0009$~c/d that is more consistent with the photometric variability. Figure~\ref{0381_rv} shows the obtained orbital solutions assuming these two periods. For the former, the amplitude is $138\pm5$~km/s, while for the latter it is slightly smaller, $122\pm6$~km/s. The corresponding binary mass functions are $f_2 = 0.0979\pm0.0102~M_{\odot}$ and $f_2 = 0.0473\pm-0.007$, respectively. In any case, this is consistent with an ELM with a cool unseen companion, as for EPIC 212682624.

Unfortunately, the mass of the primary in these systems is not straightforward to determine, due to the complicated evolution of ELMs, which involves residual burning and leads to an overlap between evolutionary tracks that show the same value of $\log~g$ for different combinations of mass, radius, and metallicity (see e.g. Fig. 9 in \citealt{2016A&A...595A..35I} and Fig. 8 in \citealt{2019ApJ...871..148L}). Therefore, further constraints on the ellipsoidal systems cannot be derived with the current set of spectra.

\subsection{Other previously known binaries}

\subsubsection{EPIC 211477347 and 212167054} These are two rather normal Algol-type eclipsing binaries
first detected by the Catalina Survey \citep{2014ApJS..213....9D}. However, they are not included
in the final catalogue of eclipsing binaries from the {\it Kepler} mission \citep{2016AJ....151...68K}.
\citet{2018ApJS..238....4P} list a period of 0.7667809\,d for EPIC 212167054.
The periods suggest that they can consist of normal (i.e. not compact) components \citep{2004A&A...417..263B}.

\subsubsection{EPIC 211617909} \citet{2009AJ....137.4011S}
included this object in their list of cataclysmic variables (CVs) but did not publish any period. Here we report a period of 0.146\,d; we also identify its harmonics. Cataclysmic variables are
binary star systems consisting of a WD and a normal star companion. They irregularly increase in 
brightness by a large factor and then drop back down to a quiescent state due to mass transfer and accretion
\citep{2020MNRAS.492L..40A}. EPIC 211617909 is included in the list of known bright WDs by \citet{2017MNRAS.472.4173R}. 
The two independent light curves show a quiet irregular behaviour, such as that of the CV variable HV And \citep{2012NewA...17..442R}.
The orbital period of EPIC 211617909 is most likely 0.146\,d. Whereas the LAMOST spectrum is very noisy, 
showing only H$\alpha$ in emission, the SDSS spectrum clearly shows that most of the hydrogen lines are in emission. Also,
the binarity characteristics are visible; this is without any doubt a CV system. The fact that it has been classified as sdA despite the emission lines is a consequence of the SDSS spectral classification pipeline allowing for negative normalization when fitting spectral templates (i.e. Balmer lines in emission can be fitted with an upsidedown A-type spectrum and be erroneously labelled as A type).

\subsubsection{EPIC 211378898} EPIC 211378898 is a known U Geminorum-type variable that has been studied for many 
years \citep{2017PASJ...69...75K}. U Geminorum-type variables are a class of dwarf novae that show long, bright 
outbursts (super-outbursts) in addition to normal outbursts, and these super-outbursts are believed to 
be caused by the tidal instability when the disk radius expands to a resonance during an outburst 
\citep{1989PASJ...41.1005O}. During super-outbursts, variations with periods slightly longer than 
the orbital period are observed, and they are called super-humps. The orbital period of
EPIC 211378898 is about 0.065\,d \citep{1996ASSL..208...61M}, which is exactly what we have measured
in the light curve. The LAMOST and SDSS spectra show the typical strong emission of the hydrogen lines. Like\ EPIC 211617909, EPIC 211378898 has been classified as sdA due to negative normalization factors being allowed.
It should be noted that this short orbital period excludes a main-sequence binary due to the Roche lobe criterion \citep[i.e. at least one of the components should be 
a compact object;][]{1983ApJ...268..368E}. 

\subsubsection{EPIC 251353301} This star is listed as a W UMa-type variable with a period of 0.3363233\,d by \citet{2014ApJS..213....9D}.
This would perfectly coincidence with the frequency of 2.9724 c/d (Table \ref{tsa_targets}). Although this is the frequency with the
lowest signal-to-noise ratio, all other detected frequencies are harmonics of it.

\subsection{Other known variable sdA stars in the AAVSO database}

\begin{table}
\begin{center}
\caption{Types of the 120 known variable stars taken from the VSX.} 
\label{vsx}
\begin{tabular}{cc}
\hline
Type & No. \\
\hline
CV & 3 \\
DSCT & 14 \\
EA & 16 \\
EB & 1 \\
EW & 37 \\
HADS & 11 \\
NON-CV & 1 \\
RRAB & 8 \\
RRC & 18 \\
RRD & 3 \\
RS & 1 \\
UGSU & 3 \\
V361HYA & 1 \\
VAR & 3 \\
\hline
\end{tabular}
\tablefoot{CV stands for cataclysmic variable. DSCT stands for variable of the $\delta$ Scuti type. EA stands for detached eclipsing binary. EW stands for W Ursae Majoris-type eclipsing binary. HADS stands for high amplitude $\delta$ Scuti stars. NON-CV stands for a star that was once classified as a CV but then was found to be constant. RRAB stands for RR Lyrae type-a or type-b variable. RRC stands for RR Lyrae type-c variable. RRAB stands for RR Lyrae type-d variable. RS stands for subtype for eclipsing or ellipsoidal systems showing chromospheric activity. UGSU stands for U Geminorum-type variable, quite often called a dwarf nova. V361HYA stands for very rapidly pulsating hot subdwarf B star. VAR stands for variable of unknown type.
}
\end{center}
\end{table}

To gather further insight into the range of variables within the sdA spectroscopic class, we cross-matched our complete list of sdA candidates with the VSX database maintained by the American Association of Variable Star Observers (AAVSO) and found 120 matches. This is only about 3\% of
the total sample. It is important to emphasize that no list of apparent non-variable stars exists, especially 
for the highly accurate space-based datasets. 
Table \ref{vsx} lists the statistics of the found variables in the VSX. The largest group are the binaries (CV, EA, EB, EW, and RS),
followed by the pulsating stars (DSCT, HADS, RRAB, RRC, RRD, and V361HYA). For the RR Lyrae stars, for example, the period-luminosity 
relation also depends on the metallicity \citep{2004ApJS..154..633C}, which could provide interesting insight into the sdA stars. 
For $\delta$ Scuti stars, there is a well-established period-luminosity-colour-metallicity relation \citep{1999A&A...352..547P}
that can be used as for RR Lyrae stars. Furthermore, for the radial $\delta$ Scuti-type pulsators of Population I, the observations indicate 
increasing and decreasing period changes, while most of the Population II counterparts are characterized by sudden jumps \citep{1998A&A...332..958B}.
These jumps have already been successfully observed \citep{2019PASP..131f4202Z}. A detailed investigation of all available
data on a longer time basis is needed to put further constraints on the evolutionary status of the pulsators. 

\section{Discussion}

Reflecting the inhomogeneity of the sdA spectral class, we find these objects to show a wide range of possible photometric variability. The K2 light curves allow us to confirm that three of the analysed objects (EPIC 202066368, 202067392, and 202066571) are consistent with main-sequence stars, as expected for most sdA stars. A fourth object (EPIC 212168575) is likely a horizontal branch star.

The rate of binaries, in particular with compact companions, seems to be high among photometric variable sdA stars. Only three out of the nine stars identified as binaries have periods long enough to not harbour a compact star. Two of the binaries (EPIC 212682624 and EPIC 246429212) have a (pre-)ELM as a primary, with an unseen cool companion as the secondary.
The derived surface gravities would be consistent with metal-poor main-sequence stars (i.e. blue stragglers).
If their detected companions are not the objects that donated mass, they could be rare examples of such objects formed in primordial (or dynamically formed) hierarchical 
triple stars \citep{2009ApJ...697.1048P}.

For the CVs, it is interesting to note that \citet{2005ASPC..334..351S} suggested that the observed 
period gap in CVs is created by a superposition of a short-period subdwarf B-type channel and a longer-period post-thermal-timescale 
mass-transfer channel rather than the conventional model of a single uniform formation channel and disrupted magnetic braking. The detected CVs among the sdA stars could
contribute to the analysis of this mechanism.

Within our analysis, about half (12 out of 25) of the objects were found to show no variability on a time basis of 30 days
and to have very low amplitudes. 
There has not yet been any homogeneous and comprehensive study on the incidence of (non-)variable
stars across the HRD. However, studying this parameter is of the upmost importance for analysing the current evolutionary status of
the sdA stars. Because of their evolutionary history across the HRD, we can expect a different behaviour than that of solar-abundant main-sequence
stars. A first basic study in this respect entitled `Gaia Data Release 2 -- Variable stars in the 
colour--absolute magnitude diagram' \citep{2019A&A...623A.110G} has been published. On the basis of the available {\it Gaia} 
observations, the authors searched for (non-)variable 
stars for which more than 20 observations in the three different bands are available. This transforms to a precision level of approximately 5 to
10\,mmag. However, the dataset they used was not published, so an analysis in respect to the sdA stars is not possible.

\section{Conclusions} \label{conclusions}

A detailed analysis of the formation and evolution of sdA stars is still in its early stages. One of the insights we have is that
the group still consists of many different types of objects, for example metal-poor A/F-type stars in the halo with overestimated 
surface gravities and extremely low-mass WDs and their precursors (i.e. ELMs and pre-ELMs).

Investigating the variable and apparently non-variable objects among this group will help to shed more light on the
astrophysical processes from a different point of view. However, the incidence of apparently non-variable stars 
is also very important because certain atmospheric conditions, such as the
degeneracy of different layers, stratification, strong magnetic fields, accretion, and stellar winds, suppress 
pulsation.

We have investigated the accurate time series of the {\it Kepler} K2 satellite mission available for 25 sdA and candidate
sdA stars. Among this sample, we have found 13 variable stars of different types. Among these are classical pulsating stars,
ELLs, eclipsing binaries, CVs, and rotationally induced variables. This mixture
of types also reflects the mentioned variety of sdA subgroups.
For the 12 apparently non-variable stars, we deduced the noise level of the amplitude spectra for different frequency
domains in order to account for the quality of the datasets.

This pilot study is just the beginning and is presented to sound the possibilities of using variable stars for the study
of sdA stars. The most recent list includes about 3900 members or candidates for this star group. Upcoming datasets, such
as from the {\it TESS} satellite, will only cover the brighter stars with an accuracy equivalent to that of the {\it Kepler}
mission. For a further study of the astrophysical parameters of the variable stars, spectroscopic observations, for example radial velocity measurements, are also needed.

\begin{acknowledgements}
This work has been supported by the DAAD (project No. 57442043) and the Erasmus+ programme of the European Union under grant number 2020-1-CZ01-KA203-078200. IP acknowledges funding by the Deutsche Forschungsgemeinschaft under grant GE2506/12-1.
It has also made use of data from the European Space Agency (ESA) mission {\it Gaia} (\url{https://www.cosmos.esa.int/gaia}), 
processed by the {\it Gaia} Data Processing and Analysis Consortium (DPAC, \url{https://www.cosmos.esa.int/web/gaia/dpac/consortium}). 
Funding for the DPAC has been provided by national institutions, in particular the institutions participating in the {\it Gaia} 
Multilateral Agreement. This paper is dedicated to Holger Pikall who died during its preparation.
\end{acknowledgements}

\bibliographystyle{aa}
\bibliography{sdA}

\end{document}